\documentclass[12pt,a4paper]{article}

\usepackage[spanish]{babel}
\usepackage[utf8x]{inputenc}
\usepackage[T1]{fontenc}
\usepackage{amsmath}
\usepackage{graphicx}
\usepackage[colorinlistoftodos]{todonotes}
\usepackage{natbib}    

\usepackage{enumerate}
\usepackage{enumitem}
\usepackage{amsmath}    
\usepackage{amssymb}    
\usepackage{bm}         
\usepackage{graphicx}   
\usepackage{color}      
\usepackage{hyperref}   
\usepackage{natbib}    
\usepackage{epsf}
\usepackage{lscape}
\bibpunct{(}{)}{;}{a}{,}{,}
\title{Una alternativa para la estimación del ingreso promedio
mediante métodos de estimación en áreas pequeñas \\\vspace{1cm}
An alternative for the average income estimation using small area methods}

\author{Julieth Casta\~neda and Cristian T\'ellez$^{a}$ and Jairo F\'uquene.$^{b}$
\\ \small $^{a}$ Department of Statistics, Santo Tomas University.
\\ \small $^{b}$ Technical Consultant for National Administrative Department of Statistics, Colombia.}

\begin{document}

\date{}
\maketitle


\renewcommand{\tablename}{Tabla}
\renewcommand{\figurename}{Figura}






\date{}


\begin{center}
\textbf{Resumen}

\end{center}

El ingreso promedio de los hogares es uno de los índices más importantes para la toma de decisiones y los modelos de desigualdad económica y pobreza. En este trabajo se propone un procedimiento práctico para estimar el ingreso promedio de los hogares mediante métodos estadísticos en área pequeñas. Se considera la encuesta multipropósito y variables auxiliares de tipo económico y demográfico tales como el índice de pobreza multidimensional, el índice de valorización y las proyecciones de población. Mediante el uso de la metodología propuesta, se encuentra que los errores estándar relativos para las estimaciones del ingreso promedio disminuyen considerablemente.

\vspace*{.3in}

\noindent \textsc{Palabras Clave}: Modelo Fay-Herriot, ingreso promedio, pobreza multidimensional, Índice de valoración, áreas pequeñas, encuesta multiproposito 2011.

\begin{center}
\textbf{Abstract}

\end{center}

The average household income is one of the most important indexes for decision making and the modelling of economic inequity and poverty.
In this work we propose a practical procedure to estimate the average income using small area methods. We illustrate our proposal
using information from a multipurpose survey and suitable economic and demographic variables
such as the multidimensional poverty and the valorization indexes and the official population projections. We find that the standard relative errors for the income average estimates improve substantially when the proposed methodology is implemented.

\vspace*{.3in}


\section{Introducci\'on}

El muestreo estadístico, a diferencia de los censos, permite obtener la información a un costo reducido. El muestreo se utiliza no solamente para la obtención de estimaciones en la población completa, sino para estimar parámetros en una variedad de subpoblaciones denominadas dominios, que se definen generalmente como áreas geográficas o grupos socioeconómicos.\\

En el contexto de la estimación en áreas pequeñas, se dice que un estimador de un parámetro en un dominio dado es directo si está basado solamente en los datos específicos del dominio. Un estimador directo puede usar también información auxiliar, como por ejemplo el total en el dominio de una variable $x$ relacionada con la variable de interés $y$. Un estimador directo es típicamente un estimador basado en el diseño muestral, aunque en ocasiones su uso se justifique con modelos. Los estimadores basados en el diseño muestral utilizan los pesos muestrales. Las inferencias derivadas de los mismos están basadas en la distribución de probabilidad inducida por el mecanismo aleatorio de extracción de la muestra, bajo el supuesto de que los valores de las variables en los elementos de la población permanecen fijos.\\

Los estimadores asistidos por modelos se introducen a partir de modelos de trabajo, pero optimizando sus propiedades de sesgo y varianza respecto de la distribución del diseño. En la literatura estadística estos estimadores también se consideran basados en el diseño muestral.\\

Un dominio es grande si la muestra específica del dominio es suficientemente grande para obtener estimadores directos con una precisión adecuada. Un dominio es pequeño en caso contrario. En este texto usaremos el término área pequeña para denotar a los dominios pequeños.\\

En este trabajo realizaremos una estimación del promedio de los ingresos de los bogotanos a nivel de localidad, para este ejercicio las localidades serán los dominios y en el contexto se SAE (Small Area Estimation) serán las áreas pequeñas que queremos estimar.\\

El trabajo está organizado de la siguiente manera: En la siguiente sección se presentan los antecedentes. En la sección 3 el marco teórico basado en el estimador directo y el modelo Fay-Herriot (\cite{fay1979estimates}, \cite{schall1991estimation}, \cite{molina2015small}), utilizados en este trabajo.  En la Sección 4 se presenta una descripción de las variables utilizadas. La Sección 5 muestra la construcción de la variable ingreso. En la Sección 6 se describe el diseño estadístico de la encuesta multipropósito. En la Sección 7 se exponen los resultados y, finalmente, en la Sección 8  se presentan las conclusiones.

\section{Antecedentes}

Colombia es una de las sociedades más desiguales a nivel mundial, algunos de los estadísticos calculados y que evidencian dicha desigualdad  son el ingreso promedio, el coeficiente de Ginni, la pobreza multidimensional, el índice de necesidades básicas insatisfechas, entre otros. Estas estadísticas  son vitales para la toma de  decisiones políticas y  la asignación de recursos en diferentes niveles de desagración geográfica. En aras de desarrollar una  metodología moderna  de estimación en este trabajo de grado se propone el cálculo del ingreso promedio en la ciudad de Bogotá  para el año 2011 por localidades, haciendo uso de estimación en áreas pequeñas y de información disponible de otros sectores.\\

El Departamento Administrativo Nacional de Estadística (DANE) realiza el cálculo del ingreso promedio en los hogares utilizando la  Encuesta Continua de Hogares (ECH) y su versión posterior la Gran Encuesta Integrada de Hogares (GEIH), además de instrumentos multipropósito orientados a  la medición de indicadores de mercado laboral.\\

La construcción de la variable ingreso  consta de varios pasos. Primero se estiman los agregados para cada una de sus fuentes. En segundo lugar se aplica un modelo de imputación para corregir sesgo en la información reportada por la población en las Encuestas de Hogares. En tercer lugar se obtiene un consolidado del ingreso total a nivel de la Población en Edad de Trabajar (PET), que más adelante se agrega a nivel de la unidad de gasto. En cuarto lugar, dependiendo del status de tenencia de la vivienda, se realiza una imputación por propiedad de la misma. Para finalizar, se calcula el valor del ingreso per cápita de la unidad de gasto, monto a partir del cual se calcula la incidencia de la pobreza y la pobreza extrema (DANE, IPM 2011).\\

En este trabajo se hará uso de la construcción de la variable Ingreso tal como es definida por  el Departamento Administrativo Nacional de Estadística (DANE), y realizaremos la estimación por localidades para Bogotá en el año 2011. Para mejorar las estimaciones se utilizan variables auxilares del mismo año. Dichas variables son el recíproco del índice de pobreza multidimensional que es calculado por  la Secretaria Distrital de Planeación (SDP), el valor comercial (índice de valorización) de los predios residenciales por localidad disponible en el Censo Inmobiliario que realiza Catastro y las proyecciones de población por localidad que para este nivel de desagregacion las publico la Secretaria Distrital de Planeación (SDP), pero están basadas en la metodología aprobada por el Departamento Administrativo Nacional de Estadística (DANE). Mediante el uso de las variables mencionadas, se aplicaran las técnicas de muestreo e inferencia finitas para calcular estimadores directos del ingreso promedio y estimar sus varianzas. Estas estimaciones directas alimentan un modelo Fay-Herriot (\cite{fay1979estimates}, \cite{schall1991estimation}, \cite{molina2015small}) (desde una perspectiva en \'areas peque\~nas, i.e., localidades) que permite calcular predictores lineales insesgados óptimos (EBLUP – empirical best linear unbiased predictor).\\

\section{Marco Teórico}

En el marco muestral de la encuesta multipropósito se considera a cada localidad como un estrato. En cada una de ellas se ordenaron las manzanas por estrato socioeconómico y usando el método sistemático se seleccionaron segmentos dentro de cada estrato. El factor final de expansión por hogar $j$, $\omega_{j}$,
es el producto de factores de expansión. El estimador directo propuesto por \cite{hajek1971comment} y el Error Estándar Relativo de este estimador para el promedio de ingresos $\bar{Y}_{d}$ pot localidad $d$ están dados por

\begin{align}
\hat{\bar{Y}}_{d}^{}&=\dfrac{\sum_{j \in s_{d}} \omega_{j} y_{j}}{\hat{N}_{d}}; &
\hat{\text{EER}}^{\text{}}(\hat{\bar{Y}}_{d})&=
\dfrac{\sqrt{\dfrac{1}{\hat{N}_{d}^{2}}\sum_{j \in s_{d}}\omega_{j}(\omega_{j}-1)
(y_{j}-\hat{\bar{Y}}_{d}^{\text{}})}}{\hat{\bar{Y}}_{d}^{\text{}}},
\label{eq:directo}
\end{align}

donde $s_{d}$ hace referencia al
conjunto muestral para la localidad $s$ y con  $\hat{N}_{d}=\sum_{j \in s_{d}} \omega_{j}$ la estimación
del tamaño poblacional en la localidad
$d=1,...,19$. El tamaño de muestra de segmentos dentro de cada localidad $d$ está dado por la siguiente fórmula:

\begin{equation}
   n_{d}=\frac{N_{d}PQdeff}{N_{d}\left ( ESrelP \right )^{2} + PQdeff},
\end{equation}
donde:

\begin{enumerate}
  \item $n_{d}$: Número de personas en la muestra dentro de la localidad $d$.
  \item $N_{d}$: Número de personas total dentro de la localidad $d$.
  \item $P$ : Porcentaje de ocurrencia de los principales indicadores. Para el cálculo del tamaño de muestra se usó fenómenos con aproximadamente una prevalencia del 10 \% a nivel de localidad.
  \item $Q =1-P$.
  \item ESrel : Error estándar relativo, que para el ejercicio se fijó en 5\%.
\end{enumerate}

El efecto del diseño, $deff= \frac{Var_{\left ( Congl \right )}} {Var_{\left ( MAS \right )}}$, es una relación, entre la varianza de los
conglomerados, $Var_{\left ( Congl \right )}$, y  la varianza bajo un diseño aleatorio simple de elementos, $Var_{\left ( MAS \right )}$, es decir, mide el efecto de los conglomerados en el diseño \cite{sarndal2003model}. Recomendaciones acerca del valor adecuado de un efecto de diseño son dadas en
Naciones \cite{onu}, página 27 recomienda utilizar un valor de efecto de diseño aproximadamente de tres unidades para la encuesta de Hogares. El modelo Fay-Herriot
\cite{fay1979estimates}
est\'a dado por

\begin{align}
\boldsymbol{Y}= \boldsymbol{X}\boldsymbol{\beta} + \boldsymbol{Z}\boldsymbol{u} + \boldsymbol{\epsilon},
\label{eq:FH}
\end{align}

donde $\boldsymbol{Y}$ es el vector que contiene
las estimaciones directas de los ingresos promedio para las localidades $d$,
$\boldsymbol{X}=\text{col}_{1\leq d \leq D}(\boldsymbol{x}_{d})$ es
la matriz de covariables,
$\boldsymbol{\beta}=(\beta_{1},...,\beta_{D})^{'}$ es el vector
de parámetros,
$\boldsymbol{Z}=I_{D}$, $\boldsymbol{u}=(u_{1},...,u_{D})^{'}$ son los efectos aleatorios para las localidades $d$ independientes e idénticamente distribuidos
donde $\boldsymbol{u} \sim \text{Normal}(\boldsymbol{u}; \boldsymbol{0},\sigma^{2}_{u}I)$, $\boldsymbol{\epsilon}=(\epsilon_{1},...,\epsilon_{D})^{'}$
es el vector de efectos aleatorios con distribución Normal $\boldsymbol{\epsilon} \sim \text{Normal}(\boldsymbol{\epsilon}; \boldsymbol{0},\sigma^{2}_{d}I)$, cuyas componentes son independientes e idénticamente distribuidas y están asociadas a las localidades.
El predictor EBLUP bajo el modelo Fay-Herriot es

\begin{align}
\hat{\bar{Y}}_{d}^{\text{EBLUP}}&=
\dfrac{\hat{\sigma}^{2}_{u}}{\hat{\sigma}^{2}_{u}+\hat{\sigma}^{2}_{d}}
\hat{\bar{Y}}_{d}^{\text{}} +
\dfrac{\hat{\sigma}^{2}_{d}}{\hat{\sigma}^{2}_{u}+\hat{\sigma}^{2}_{d}}
\boldsymbol{x}_{d}\hat{\boldsymbol{\beta}}.
\label{eq:EBLUP}
\end{align}

Donde $\hat{\bar{Y}}_{d}$ es el estimador directo y $\boldsymbol{x}_{d}\hat{\boldsymbol{\beta}}$ es el estimador sintético.
El estimador (\ref{eq:EBLUP}) es una combinación convexa
del estimador directo y el estimador sintético en donde las varianzas
de los efectos aleatorios $\hat{\sigma}^{2}_{u}$ y
del estimador directo $\hat{\sigma}^{2}_{d}$ juegan un papel
importante en los pesos. Por ejemplo, cuando $\hat{\sigma}^{2}_{u}$ es
pequeña en comparación con
$\hat{\sigma}^{2}_{d}$, se tiene un mayor peso en la estimaciones
sintéticas. Por el contrario, cuando $\hat{\sigma}^{2}_{d}$  es
pequeña en comparación con $\hat{\sigma}^{2}_{u}$,  se tiene un mayor peso
en las estimaciones directas.\\

Para la estimación del coeficiente de variación
se utiliza una aproximación del error cuadrático medio
del estimador $\hat{\bar{y}}_{d}^{\text{FH}}$ \cite{prasad1990estimation}.

\begin{align}
\text{MSE}(\hat{\bar{Y}}_{d}^{\text{EBLUP}})=
f_{1}(\hat{\sigma}^{2}_{u})+
f_{2}(\hat{\sigma}^{2}_{u})+
fg_{3}(\hat{\sigma}^{2}_{u}),
\label{eq:MSE}
\end{align}

donde

\begin{align*}
f_{1}(\hat{\sigma}^{2}_{u})&=\dfrac{\hat{\sigma}^{2}_{u}\hat{\sigma}^{2}_{d}}{
\hat{\sigma}^{2}_{u}+\hat{\sigma}^{2}_{d}}; &
f_{2}(\hat{\sigma}^{2}_{u})&=\dfrac{\hat{\sigma}^{4}_{d}}{
(\hat{\sigma}^{2}_{u}+\hat{\sigma}^{2}_{d})^{2}}(\boldsymbol{X}^{'}\hat{\boldsymbol{V}}^{-1}\boldsymbol{X})^{-1}\boldsymbol{x}_{d}^{'}; &
f_{3}(\hat{\sigma}^{2}_{u})&=\dfrac{\hat{\sigma}^{4}_{d}}{
(\hat{\sigma}^{2}_{u}+\hat{\sigma}^{2}_{d})^{3}}\text{avar}(\hat{\sigma}^{2}_{u}),
\end{align*}

donde

\begin{align*}
\text{avar}(\hat{\sigma}^{2}_{u})=2\left(\sum_{d=1}^{D}
1/(\hat{\sigma}^{2}_{u}+\hat{\sigma}^{2}_{d})^{2}
\right)^{-1}.
\end{align*}

\section{Variable de interés e información auxiliar}

En esta sección se describen las variables
auxiliares utilizadas en el modelo Fay-Herriot y se describe el cálculo de la variable de interés, posteriormente las variables auxiliares, La primera  está relacionada con el índice de pobreza multidimensional y la segunda con el índice de valorización de los predios.  Como se menciona en la sección de resultados otra variable auxiliar utilizada son las proyecciones de población para las localidades de Bogotá con el objetivo de mejorar el modelamiento de las variables auxiliares propuestas. Con respecto a encuestas por muestreo en Latinoamérica, en  \cite{molina2018desagregacion} se encuentra una guía completa del uso de métodos de estimación en áreas pequeñas mediante la encuesta de hogares.

\subsection{Encuesta Multipropósito}

La Encuesta Multipropósito para Bogotá - Distrito Capital año 2011 \cite{bustamante2011lanzamiento} surge como una necesidad del Distrito de obtener información socioeconómica periódica para la formulación, seguimiento y evaluación de las políticas distritales dando así continuidad a las encuestas de Calidad de Vida de los años 1991, 1993, 2003 y 2007 y la Encuesta de Capacidad de Pago 2004, en el marco de unificación de sus contenidos temáticos. Estos contenidos profundizan en aspectos de cobertura, calidad y gasto de los hogares en servicios públicos domiciliarios, mercado laboral y condiciones de vida. En este sentido, la Secretaría Distrital de Planeación (SDP) consideró conveniente integrar la temática de las dos encuestas en una Encuesta Multipropósito, manteniendo los niveles de desagregación por localidad y estrato.\\

Adicionalmente, el diseño de esta encuesta incluye temas en torno a problemas específicos que afectan la calidad de vida urbana y la capacidad de pago de los hogares de Bogotá, permitiendo comparabilidad con otras ciudades del país y a nivel internacional. De otra parte, se busca una aplicación bianual de dicha encuesta, con el fin de realizar el seguimiento permanente a las necesidades de la ciudad. De tal forma se podr\'ia permitir definir una línea base para los planes de desarrollo futuros, y su evaluación a la mitad y al final de cada administración. Algunos usos importantes de métodos de estimación han sido estudiados en \cite{gutierrez2015estimacion} donde el objetivo principal es el cálculo por muestreo del coeficiente de  Ginni.\\

De la encuesta Multipropósito  se usan  las variables que contienen todas las posibles fuentes de ingresos que tienen las personas, es decir, si son empleados, independientes o reciben algún tipo de ingreso por una actividad económica, las variables son las siguientes:

\begin{itemize}

\item K30: ¿Cuánto ganó el mes pasado en este empleo (incluya propinas y comisiones y excluya viáticos y pagos en especie)?
\item K35: ¿El mes pasado recibió subsidio de alimentación en dinero?
\item K36: ¿El mes pasado recibió auxilio de transporte en dinero?
\item K37: ¿El mes pasado recibió subsidio familiar en dinero?
\item K38: ¿El mes pasado recibió primas (técnica, de antigüedad, clima, orden público, etc.) en dinero?
\item K39A: ¿Durante los últimos doce meses recibió prima de servicios?
\item K39B: ¿Durante los últimos doce meses recibió prima de navidad?
\item K39C: ¿Durante los últimos doce meses recibió prima de vacaciones?
\item K39D: ¿Durante los últimos doce meses recibió bonicaciones?
\item K39E: ¿Durante los últimos doce meses recibió pagos o indemnizaciones por accidentes de trabajo?
\item K40: ¿Cuál fue la ganancia neta o los honorarios netos de ... en esa actividad, negocio, profesión o finca, el mes pasado?
\item K46: ¿El mes pasado tuvo otros trabajos o negocios por los cuales recibió ingresos?
\item K51: ¿El mes pasado recibió algún ingreso por concepto de pensión de jubilación, sustitución pensional, invalidez o vejez?
\item K52: ¿El mes pasado recibió algún ingreso en dinero para el sostenimiento de hijos menores de dieciocho años (incluya pensión de alimentación y contribución de padres ausentes)?
\item K53: ¿El mes pasado recibió algún ingreso por concepto de arriendos de casas, apartamentos, fincas de recreo, lotes, vehículos, maquinaria y equipo?
\item K54: ¿Durante los últimos doce meses recibió primas por pensión de jubilación o por sustitución pensional?
\item K55: ¿Durante los últimos doce meses recibió algún ingreso por concepto de ayudas en dinero provenientes de otros hogares o instituciones (padres, hijos, familiares, amigos)?
\item K56: Durante los últimos doce meses, ¿ ... recibió
dinero por venta de propiedades (casas,
edificios, lotes, maquinaria, vehículos,
electrodomésticos, etc.)?
\item K57: ¿Durante los últimos doce meses recibió dinero por otros conceptos (cesantías, intereses por cesantías, intereses por préstamos o CDT, rifas, etc.)?

\end{itemize}

\subsection{Índice de Pobreza Multidimensional}

En esta sección se describe el procedimiento básico para el cálculo de indicador de pobreza multidimensional \cite{salazar2011indice}. Para la medición de la pobreza multidimensional se requiere inicialmente determinar la unidad de análisis, las dimensiones, las variables o indicadores contemplados para cada una de las dimensiones así como también  los pesos de cada dimensión y de cada variable. Adicionalmente, es necesario escoger las líneas de
pobreza para cada indicador con las cuales se puede identificar quién o qué hogar es pobre bajo el enfoque multidimensional.\\

\begin{figure}[ht]
\centering
\includegraphics[width=0.95\textwidth]{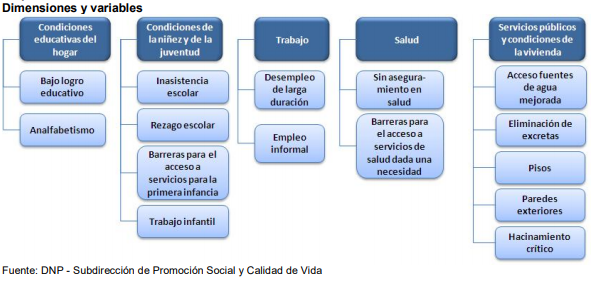}
\caption{Diagrama de flujo para el cálculo del índice de pobreza multidimensional para Bogotá.}
\label{fig:figura1}
\end{figure}

El diagrama de flujo presentado en la Figura 1  presenta las dimensiones y las variables consideradas para el cálculo del índice de pobreza multidimensional para Bogotá. Una vez definidos estos aspectos, el cálculo del indicador comienza con la evaluación de los desempeños de los individuos en cada variable para identificar los hogares que se encuentran en privación al presentar desempeños por debajo de las líneas de pobreza definidas previamente.\\

Por otro lado, para clasificar un hogar con pobreza multidimensional se requiere que presente carencia o privación de varios aspectos simultáneamente. Esto ocurre cuando los indicadores ponderados en los cuales sufre privación, suman al menos el 30 $\%$.
Una vez identificados los hogares pobres, se realiza el proceso de agregación basado en el sistema de medición tradicional \cite{foster2010foster}, que permite obtener entre otros: 1) el porcentaje de hogares pobres, 2) Incidencia  y 3) la proporción media de dimensiones en los cuales los hogares son
privados – Intensidad –. Finalmente, el IPM también denominado como la tasa de recuento ajustada, se calcula como el producto de la incidencia y de la intensidad.\\

La primer variable auxiliar utilizada en el modelo Fay-Herriot es el reciproco de la incidencia y como se mostrará en la sección de resultados, es razonable suavizar este reciproco para obtener una relación lineal con el estimador directo de los ingresos.

\subsection{Censo inmobiliario e índice de valorización}

El catastro es el “inventario o censo, debidamente actualizado y clasificado, de los bienes inmuebles pertenecientes al Estado y a los particulares, con el objeto de lograr su correcta identificación física, jurídica, fiscal y económica” (Resolución 070 de 2011 del IGAC). Este tipo de censos podrían dar
una perspectiva al crecimiento urbano
\cite{barrera2008dinamica}.
La Unidad Administrativa Especial de Catastro Distrital, es la entidad encargada de realizar el censo catastral y su misión es la de actualizar el inventario predial de los bienes inmuebles ubicados en el Distrito Capital de Bogotá, determinando sus características físicas, jurídicas y económicas, con una excelente calidad que satisfaga las necesidades y expectativas tanto del sector público como privado teniendo en cuenta las normas de carácter Nacional y Distrital.\\

Anualmente (desde el 2010) la entidad realiza el Censo Inmobiliario de Bogotá o proceso de actualización catastral, que consiste en la revisión de los elementos físicos, jurídicos y económicos de los predios. \\

En el aspecto físico se identifican los linderos del terreno y las edificaciones existentes; en el elemento jurídico el registro de la relación que hay entre el inmueble y el propietario y/o  poseedor; y la parte económica, es la fijación del avalúo catastral mediante investigación y análisis del mercado inmobiliario. Este proceso está regulado por la Ley 14 de 1983, el Decreto 3496 de 1983 Reglamentario de la Ley 44 de 1990 y la Resolución 2555 de 1988. En el año 2010 se realizó la actualización  de 2.181.000 predios urbanos, información que quedó vigente a partir del 1 de enero del año 2011.\\

Es importante agregar que la Actualización Catastral consiste en el conjunto de operaciones destinadas a renovar los datos de la formación catastral, revisando los elementos físico y jurídico del catastro, eliminando en el elemento económico las disparidades originadas por cambios físicos, variaciones de uso o de productividad, obras públicas, o condiciones locales del mercado inmobiliario.\\
Esta Actualización se hace a través de varias etapas:\\

\begin{enumerate}
  \item La primera es de Pre-reconocimiento: donde se verifica la información que reposa en las bases de datos de la Catastro Bogotá, con la realidad física del inmueble.
  \item El siguiente paso es hacer el Reconocimiento predial, es decir, que se ingresa a los predios desactualizados y se miden las nuevas construcciones o demoliciones, se asigna un uso a esas construcciones, se asigna un destino al  predio y se califican las características de las construcciones.
  \item Finalmente se realiza la Actualización Jurídica, donde se cruza la base de datos de la Unidad con la información de la Oficina de Registro de Instrumentos Públicos.
De manera adicional, existen otras actividades del proceso de actualización catastral, que también son importantes, como el Control de Calidad (validaciones preliminares, acompañamiento en campo, muestra estadística y validaciones finales);  Análisis de fuentes secundarias: (EAAB, CODENSA, HABITAT, DANE, Curadurías) y finalmente se realizan varios análisis a dicha información.
\end{enumerate}

La segunda variable auxiliar utilizada en el
modelo Fay-Herriot proviene del índice de valorización obtenido a partir del censo inmobiliarion, en la sección de resultados se mostrará que es razonable ajustar un modelo con esta variable y las proyecciones de población y  utilizar los residuales del ajuste de estas dos variables, debido a que se obtiene una relación lineal con el estimador directo de los ingresos.

\subsection{Proyecciones de población}

El Departamento Administrativo Nacional de Estadística (DANE), en función de su papel
como coordinador del Sistema Estadístico Nacional (SEN) y en el marco del proyecto de
Planificación y Armonización Estadística, trabaja para el fortalecimiento y consolidación
del SEN, mediante la producción de estadísticas estratégicas; la generación, adaptación,
adopción y difusión de estándares; la consolidación y armonización de la información; la
articulación de instrumentos, actores, iniciativas y productos, para mejorar la calidad,
disponibilidad, oportunidad y accesibilidad, como respuesta a la demanda cada vez mayor
de información estadística estratégica.\\

Uno de los objetivos del Departamento Administrativo Nacional de Estadística (DANE), es Producir información acerca de los cambios esperados en el crecimiento, tamaño, composición y distribución de la población, a partir de los supuestos sobre la probable evolución de las componentes de la dinámica de la población – natalidad, mortalidad y migración, insumos demográficos básicos para la planeación y gestión del desarrollo nacional y territorial.\\

Las proyecciones de población se elaboran, a partir de la aplicación de modelos específicos, y en
función del nivel de especificidad requerido, de la información, la tecnología y los recursos
disponibles. Para ello se plantean diferentes escenarios acerca del comportamiento esperado de la
evolución de los componentes de la dinámica poblacional, apoyados en información del presente y
del pasado reciente teniendo en cuenta a su vez las estrategias gubernamentales y de los
particulares respecto a los programas y proyectos orientados al desarrollo, así como las
condiciones demográficas particulares de la población a los diferentes niveles administrativos del
territorio nacional.\\

Como fundamentos teóricos principales de los modelos y los métodos empleados para la
elaboración de las proyecciones de población, se siguen los principios de la teoría de la Transición
Demográfica, que hacen referencia al cambio que experimentan las poblaciones humanas al pasar
de altas a bajas tasas de natalidad y mortalidad , en algún momento de su historia.\\

Las proyecciones de población por localidad para Bogotá año 2011 que  que usaremos en este trabajo son un trabajo conjunto del  Departamento Administrativo Nacional de Estadística (DANE) y de la Secretaria Distrital de Planeación y serán usadas para ajustar un modelo de regresión lineal con el indice de valorización catastral, los residuales de este modelo se usarán como variable auxiliar en el modelo Fay-Herriot que se ajustara y los resultados se encuentran en la sección 7.

\section{Conformación de la variable Ingreso}

 El propósito es construir ingreso para todos y cada uno de los perceptores que conforman la Población en Edad de Trabajar (PET), teniendo en cuenta las diferencias entre los diversos grupos que la componen. Principalmente la desagregación entre la Población Económicamente Inactiva (PEI) y la Población Económicamente Activa (PEA), y la división de cada una de estas categorías en otros grupos.\\

Para efectos de la construcción del ingreso los individuos que conforman la PET se agrupan en cuatro categorías de perceptores de ingreso: asalariado, independiente (cuenta propia y patronos),
trabajadores familiares sin remuneración y desocupados e inactivos. El ingreso individual correspondiente a cada una de estas categorías está compuesto por al menos uno de los siguientes cinco
tipos de ingreso:

\begin{itemize}
    \item Ingreso monetario primera actividad (IMPA). El ingreso monetario mensual de los asalariados correspondiente
a sueldos y salarios, subsidios, horas extras, bonificaciones y viáticos.  Para los independientes incluye ganancia neta u honorarios primera actividad.

\item  Ingreso en especie (IE). Solo se aplica para asalariados e incluye: alimentos, vivienda, transporte, y otros como bonos sodexo y/o electrodomésticos.

\item Ingreso segunda actividad (ISA). Se aplica para todos los ocupados (asalariados, independientes y trabajadores familiares sin remuneración) e incluye: ingreso en dinero y/o en especie.

\item Ingreso monetario de desocupados e inactivos (IMDI). Incluye el ingreso por trabajo de desocupados e
inactivos realizado en periodos anteriores al de referencia.

\item Ingresos por otras fuentes (IOF). Conformado por arriendos (alquileres efectivos); intereses y dividendos
por inversiones; pensiones o jubilaciones por vejez, invalidez o sustitución pensional; ayudas7 (de hogares dentro y fuera del país, y de instituciones), pensión alimenticia por paternidad, divorcio o separación; ganancias ocasionales y cesantías e intereses por cesantías
\end{itemize}

Teniendo en cuenta lo anterior, de la encuesta multipropósito año 2011, \cite{bustamante2011lanzamiento} se usa el capítulo k Fuerza de trabajo, y se construye esta variable.

\section{Diseño Estadístico}
\subsection{Tipo de operación estadística}
La Encuesta Multipropósito Bogotá – EMB- es una encuesta por muestreo
probabilístico dirigida a hogares con entrevista cara a cara e informante directo.
\subsection{Universo}
El universo para la EMB está compuesto por los hogares particulares y la población
civil no institucional existente en el año 2011 en la parte urbana del distrito capital.
\subsection{Población objetivo}
La población objetivo está compuesta por los hogares particulares y población civil no
institucional existente en el año 2011 en la parte urbana del distrito capital en donde se
excluye lo siguiente:
\begin{itemize}
\item Hogares ubicados en las zonas rurales de la ciudad en Usaquén, Chapinero,
Santafé, San Cristóbal, Usme, Suba, Ciudad Bolívar y Sumapaz.
\item Cárceles o centros de rehabilitación penitenciarios, orfanatos o albergues infantiles, hogares geriátricos o asilos de ancianos, conventos, seminarios o monasterios, internados de estudio, cuarteles guarniciones o estaciones de policía, campamentos de trabajo, albergues para desplazados y reinsertados, centros de rehabilitación no penitenciarios, ni unidades económicas o agropecuarias
\end{itemize}
\subsection{Cobertura y desagregación geográfica}
Para el caso de la Encuesta Multipropósito Bogotá, la cobertura geográfica es la cabecera de Bogotá Urbana (19 localidades) desagregada por localidad y estrato socioeconómico.
\subsection{Unidad de observación}
Las unidades de observación son los hogares y las personas que los conforman, al igual que las viviendas que habitan ubicadas dentro de un determinado predio. A cada predio se le asocian todas las viviendas, hogares y personas que lo conforman.
\subsection{Periodo de recolección}
El periodo de recolección de la información comprende los días 7 de Febrero hasta 7
de abril de 2011.
\subsection{Marco muestral}
El marco muestral está constituido por el inventario cartográfico y el listado de viviendas y hogares a nivel de manzana, obtenidos de la información del Censo Nacional de Población y Vivienda de 2005 para la ciudad de Bogotá. Este marco se encuentra asociado al Código Homologado para Información Predial (CHIP) que identifica los predios a partir de la base de datos predial Catastral con corte a primero de marzo de 2010, suministrada por la Secretaría Distrital de Planeación – SDP-. El marco cuenta con 1.307.562 registros de predios urbanos con algún uso habitacional ubicados en 36.383 manzanas de las 19 localidades de Bogotá D.C; Usaquén, Chapinero, Santafé, San Cristóbal, Usme, Tunjuelito, Bosa, Kennedy, Fontibón, Engativá, Suba, Barrios Unidos, Teusaquillo, Los Mártires, Antonio Nariño, Puente Aranda, La Candelaria, Rafael Uribe y Ciudad Bolívar.
\subsection{Diseño Muestral}
Teniendo en cuenta los objetivos de la EMB y las consideraciones anteriores, se optópor un diseño probabilístico, estratificado, de conglomerados.
\begin{itemize}
\item Probabilístico
\end{itemize}
Cada unidad de muestreo tiene una probabilidad de inclusión conocida y mayor que cero. Esta característica del diseño es fundamental para la aplicación de la teoría de inferencia estadística, es decir, los resultados obtenidos de esta encuesta se pueden inferir a la población objetivo, teniendo en cuenta el error estándar relativo.
\begin{itemize}
\item Estratificado
\end{itemize}
Cada manzana del marco muestral se clasificó en un sólo estrato socioeconómico, acorde a la estratificación definida por la SDP. Si en el registro predial de la SDP, una manzana posee más de un estrato, a esa manzana se le asignó el estrato más frecuente de los predios que la componen. De esta forma, la estratificación del diseño se hizo con respecto al estrato socioeconómico y a la localidad que pertenece cada segmento.
\begin{itemize}
\item Conglomerados
\end{itemize}
Un conglomerado corresponde a un conjunto de predios ubicados dentro de la misma manzana o manzanas cercanas, a este conjunto de predios se le denomina segmento. En cada segmento seleccionado, se encuestan todas las viviendas, todos los hogares y todas las personas que los conforman. Por estudios de relación costo-precisión anteriores, se ha encontrado que los segmentos deben ser construidos con un tamaño de 8 predios en promedio.

\section{Resultados}

\begin{figure}[ht]
\begin{center}
\begin{tabular}{ccc}
\includegraphics[scale=0.55]{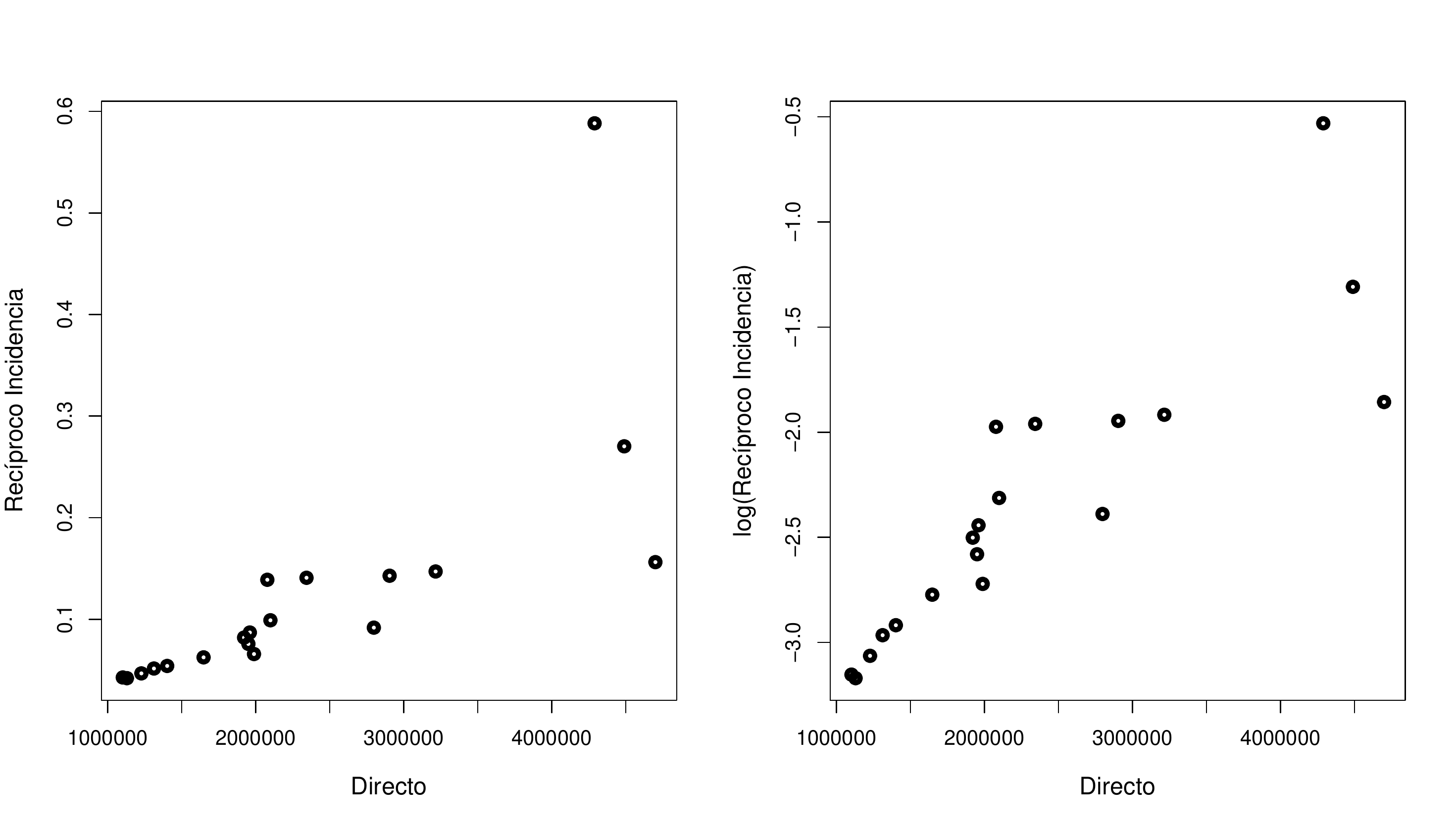}
\end{tabular}
\end{center}
\caption{Estimador directo Vs recipro de incidencia del índice de pobreza y el estimador directo Vs logaritmo natural del recíproco de incidencia del índice de pobreza }
\label{fig:mt2}
\end{figure}

Con el objetivo de mejorar la estimación del ingreso promedio por localidad se utilizan variables auxiliares de diferentes fuentes
de informaci\'on  en Colombia como son Planeaci\'on Distrital y Catastro que contienen informaci\'on econ\'omica importante
sobre las localidades en Bogot\'a, se puede observar en la Figura \ref{fig:mt2} tomando el logaritmo natural del recíproco de la incidencia del índice de pobreza multidimensional se obtiene una relaci\'on lineal positiva.

Se observa en la Figura \ref{fig:mt2} que no hay una relación lineal considerable entre el estimador directo Vs recípro del índice de incidencia de la pobreza multidimensional  no muestran empíricamente una relación lineal por lo tanto, se utiliza el logaritmo natural recípro del índice de incidencia de la pobreza multidimensional y al realizar el gráfico, se observa una relación lineal.

\begin{figure}[ht]
\begin{center}
\begin{tabular}{ccc}
\includegraphics[scale=0.55]{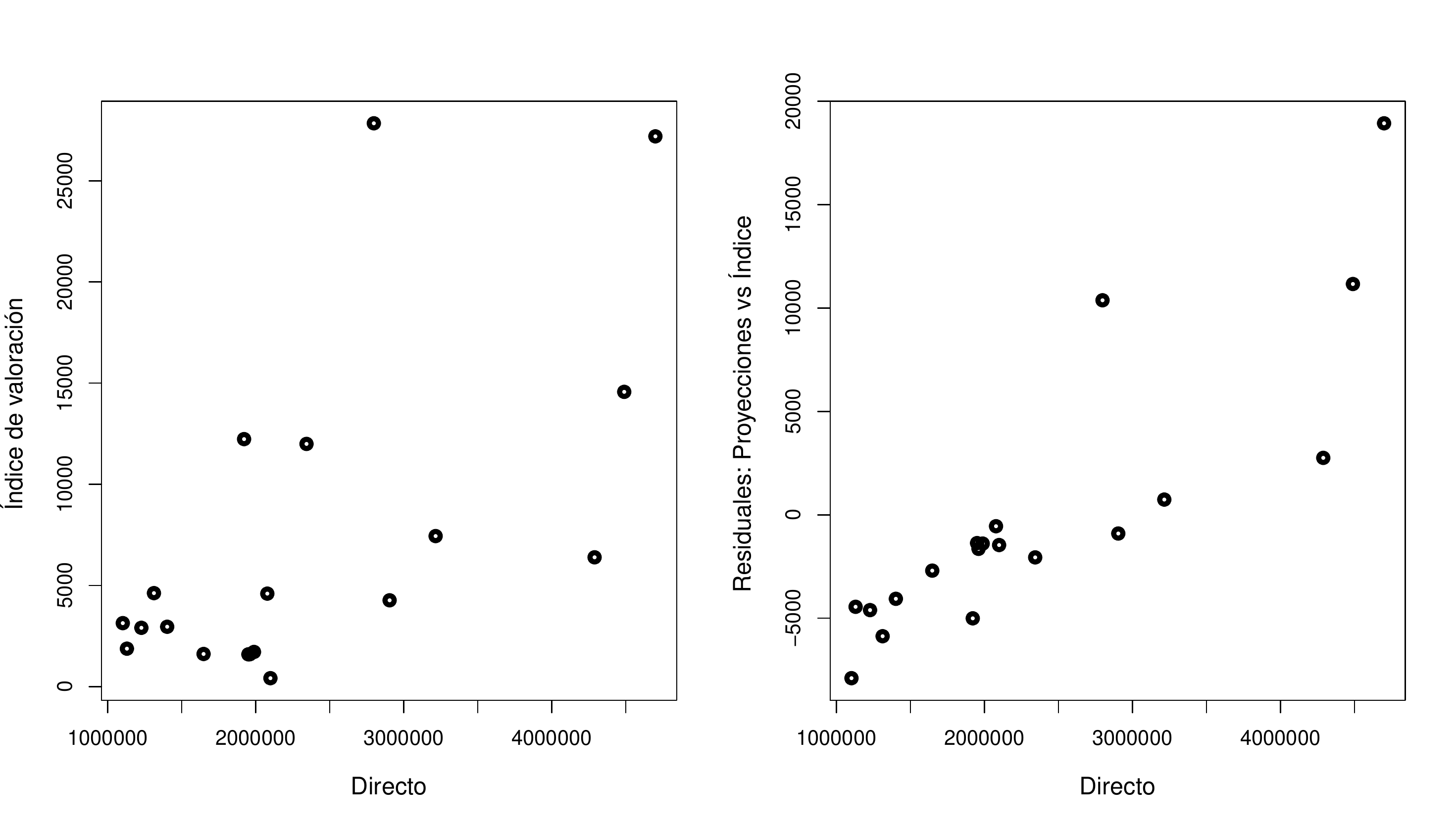}
\end{tabular}
\end{center}

\caption{Estimador directo Vs índice de valoración y Estimador directo Vs los residuale del modelo ajustado por el indice de valorización y las proyecciones de población.}

\label{fig:mt3}
\end{figure}

Por otra parte como se ilustra en la Figura \ref{fig:mt3} la variable \'indice de valoraci\'on ($\log(RI)_{d}$)
no tiene una relaci\'on lineal considerable con respecto al estimador directo. Sin embargo, si se ajusta un modelo de regresi\'on simple
usando como variable dependiente el \'indice de valoraci\'on
con respecto a la variable independiente
(proyecciones de poblaci\'on), los residuales ($\zeta_{d}$) obtenidos muestran una relaci\'on lineal con el estimador directo. Tomar los residuales como variable auxiliar ha sido estudiado por ejemplo en \cite{pratesi2016analysis}, en donde se muestra un análisis extenso de estimación en áreas pequeñas del índice de pobreza.  En algunos de los modelos estudiados en \cite{pratesi2016analysis}, las variables auxiliares son obtenidas mediante los residuales de modelos ajustados de variables obtenidos por un Censo. Para este trabajo resulta coherente, desde un punto de vista práctico, utilizar los residuales del modelo con  las variables \'indice de valoraci\'on y proyecciones de población debido a su relación lineal en las localidades estudiadas.\\

Por tanto, las variables auxiliares
consideradas son el logaritmo natural del recíproco de la incidencia del índice de pobreza ($\log(RI)_{d}$) y los residuales del modelo ajustado entre el índice de valorización y las proyecciones de población, $\zeta_{d}$. Los modelos con intercepto
considerados contienen las siguientes variables dependientes:
\begin{enumerate}
  \item Modelo 1: $$
\hat{\bar{Y}}_{d}^{}=\beta_{0}+\beta_{1}\log(RI)_{d}  + u_{d}+\epsilon_{d}.
$$
  \item Modelo 2: $$
\hat{\bar{Y}}_{d}^{}=\beta_{0}+\beta_{2}\zeta_{d}+ u_{d} + \epsilon_{d}.
$$
  \item Modelo 3: $$
\hat{\bar{Y}}_{d}^{}=\beta_{0}+\beta_{1}\log(RI)_{d}+\beta_{2}\zeta_{d}+ u_{d} + \epsilon_{d}.
$$
  \item Modelo 4: $$
\hat{\bar{Y}}_{d}^{}=\beta_{0}+\beta_{1}\log(RI)_{d}+\beta_{2}\zeta_{d}
+\beta_{3}\log(RI)_{d}\zeta_{d} + u_{d} + \epsilon_{d}.
$$
\end{enumerate}
donde $d=1,...,D$.

{\renewcommand{\arraystretch}{1.5}
\begin{table}[ht]
\begin{center}
\small
\begin{tabular}{rlrrrrrrr}  \hline   \hline
   Localidad &
  $\hat{\text{EER}}^{\text{}}(\hat{\bar{Y}}_{d}^{\text{EBLUP}_{1}})$ &
  $\hat{\text{EER}}^{\text{}}(\hat{\bar{Y}}_{d}^{\text{EBLUP}_{2}})$ &  $\hat{\text{EER}}^{\text{}}(\hat{\bar{Y}}_{d}^{\text{EBLUP}_{3}})$ &  $\hat{\text{EER}}^{\text{}}(\hat{\bar{Y}}_{d}^{\text{EBLUP}_{4}})$ \\
  \hline   \hline
Usaquén
 & 6.12 & 5.87 & 5.24 & 5.12 \\
Chapinero
 & 5.06 & 5.13 & 4.20 & 4.46 \\
Santafé
 & 4.97 & 4.96 & 4.78 & 4.79 \\
San Cristóbal
& 4.63 & 4.62 & 4.56 & 4.54 \\
Usme
 & 4.75 & 4.73 & 4.69 & 4.67 \\
Tunjuelito
 & 3.42 & 3.42 & 3.37 & 3.37 \\
Bosa
 & 5.47 & 5.49 & 5.40 & 5.35 \\
Kennedy
 & 3.37 & 3.40 & 3.35 & 3.36 \\
Fontibón
 & 4.48 & 4.56 & 4.18 & 4.15 \\
Engativá
 & 3.59 & 3.65 & 3.47 & 3.48 \\
Suba
 & 4.28 & 4.20 & 4.02 & 4.14 \\
Barrios Unidos
 & 4.17 & 4.25 & 3.98 & 3.99 \\
 Teusaquillo
 & 4.36 & 4.61 & 4.13 & 4.07 \\
 Los Mártires
 & 5.18 & 5.20 & 4.89 & 4.88 \\
 Antonio Nariño
 & 4.27 & 4.30 & 4.09 & 4.09 \\
 Puente Aranda
 & 3.70 & 3.75 & 3.51 & 3.49 \\
 Candelaria
 & 5.65 & 5.75 & 5.17 & 5.13 \\
 Rafael Uribe
 & 4.87 & 4.87 & 4.76 & 4.74 \\
Ciudad Bolívar
 & 5.81 & 5.82 & 5.85 & 5.82 \\
  \hline   \hline
\end{tabular}
\caption{Errores estándar relativos para los Modelos 1 a 4.}
\end{center}
\label{tab:mm1}
\end{table}
}

{\renewcommand{\arraystretch}{1.5}
\begin{table}[ht]
\begin{center}
\small
\begin{tabular}{rlrrrrrrr}  \hline   \hline
  & Localidad &  & $\hat{\bar{Y}}_{d}^{}$ & $\hat{\text{EER}}^{\text{}}(\hat{\bar{Y}}_{d})$ & $\hat{\bar{Y}}_{d}^{\text{EBLUP}_{4}}$ & $\hat{\text{EER}}^{\text{}}(\hat{\bar{Y}}_{d}^{\text{EBLUP}_{4}})$  & $\delta_{d}$(\%) & $n_{d}$\\
  \hline   \hline
 & Usaquén &  & 4700041.98 & 6.21 & 4522691.09 & 5.12 & 17.55 & 774.00 \\
   & Chapinero &  & 4489429.67 & 5.27 & 4508718.79 & 4.46 & 15.37 & 744.00 \\
   & Santafé &  & 1988468.69 & 5.02 & 1931813.13 & 4.79 & 4.58 & 835.00 \\
   & San Cristóbal &  & 1227721.93 & 4.65 & 1235745.29 & 4.54 & 2.37 & 941.00 \\
   & Usme &  & 1129973.22 & 4.77 & 1138877.01 & 4.67 & 2.10 & 1023.00 \\
   & Tunjuelito &  & 1647922.41 & 3.44 & 1648509.96 & 3.37 & 2.03 & 1057.00 \\
   & Bosa &  & 1312577.05 & 5.53 & 1313435.92 & 5.35 & 3.25 & 1116.00 \\
   & Kennedy &  & 1921478.60 & 3.40 & 1896320.11 & 3.36 & 1.18 & 828.00 \\
   & Fontibón &  & 3215374.29 & 4.59 & 2997159.81 & 4.15 & 9.59 & 738.00 \\
   & Engativá &  & 2343323.01 & 3.67 & 2363200.78 & 3.48 & 5.18 & 833.00 \\
   & Suba &  & 2798072.41 & 4.32 & 2850246.44 & 4.14 & 4.17 & 857.00 \\
   & Barrios Unidos &  & 2904528.79 & 4.27 & 2776390.29 & 3.99 & 6.56 & 862.00 \\
   & Teusaquillo &  & 4289301.98 & 4.59 & 4368820.90 & 4.07 & 11.33 & 789.00 \\
   & Los
Mártires &  & 1950547.15 & 5.28 & 1939882.86 & 4.88 & 7.58 & 721.00 \\
   & Antonio Nariño &  & 1960456.21 & 4.34 & 1973691.88 & 4.09 & 5.76 & 829.00 \\
   & Puente Aranda &  & 2078521.53 & 3.78 & 2167315.42 & 3.49 & 7.67 & 944.00 \\
   & Candelaria &  & 2099526.91 & 5.85 & 2120417.97 & 5.13 & 12.31 & 709.00 \\
   & Rafael Uribe &  & 1401987.26 & 4.91 & 1409329.45 & 4.74 & 3.46 & 1032.00 \\
   & Ciudad Bolívar &  & 1102178.44 & 5.85 & 1099417.97 & 5.82 & 0.51 & 876.00 \\
  \hline   \hline
\end{tabular}
\caption{Resultados obtenidos mediante el estimador directo y el EBLUP del modelo 4
con los correspondientes  errores estándar relativos y la diferencia relativa en porcentaje.}
\end{center}
\label{tab:mm2}
\end{table}
}

La Tabla 1 muestra los resultados de todos los modelos en donde el Modelo 4 presenta, en la mayoría de las localidades, el menor Error Estándar Relativo. Por tanto, tomando como criterio el cociente entre la raíz del error cuadrático medio y el promedio de ingresos (error estándar relativo o cve), el Modelo 4 es el más adecuado en términos de precisión.

Por otro lado, con el objetivo de medir el ajuste de los modelos Fay-Herriot mediante un método de selección conocido, considerados  el Akaike Information Criterion (AIC). Para los modelos considerados el AIC
est\'a dado por 557.14, 562.61, 530.52 y 528.22 (para los modelos 1 a 4 respectivamente). Por tanto, el modelo que presenta el mejor ajuste
es el modelo con una interacci\'on (modelo 4, si se toma como criterio de evaluacion el AIC).

\begin{figure}[ht]
\begin{center}
\begin{tabular}{ccc}
\includegraphics[scale=0.95]{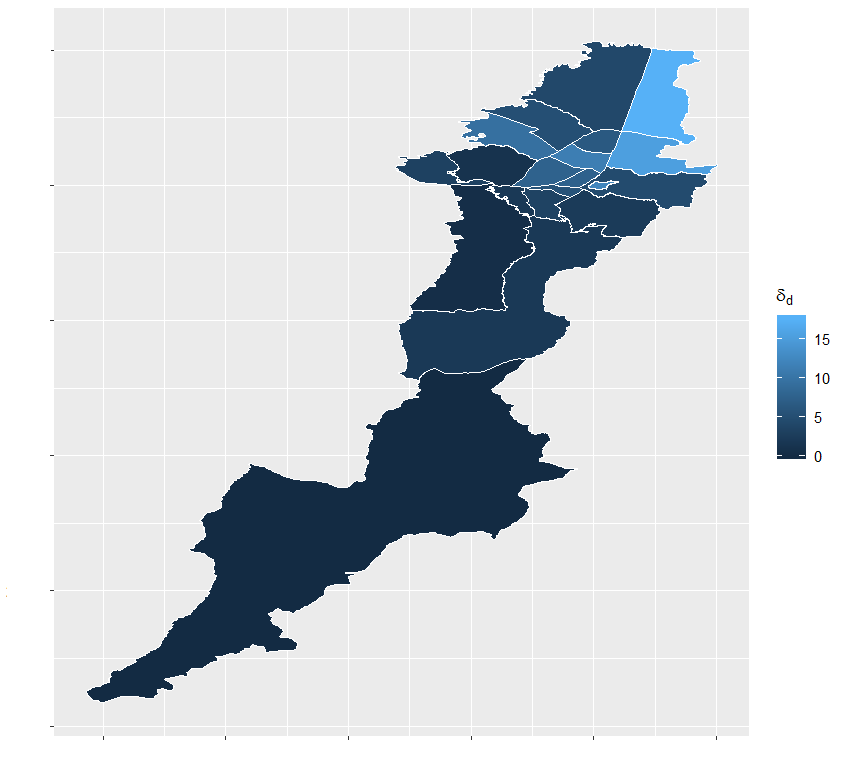}
\end{tabular}
\end{center}
\caption{Diferencia relativa entre los EER mediante el estimador directo y el modelo Fay Herriot con interacción (Modelo 4) para las diferentes localidades.}
\label{fig:mt4}
\end{figure}

La Tabla 2 muestra los resultados
obtenidos y la comparaci\'on del estimador directo con respecto a la aplicaci\'on del modelo Fay Herriot con interacci\'on. Tambi\'en
se calcula, como medida de comparaci\'on, la diferencia relativa entre los respectivos errores estándar relativos (EER): $$\delta_{d}=(
\hat{\text{EER}}^{\text{}}(\hat{\bar{Y}}_{d})-
\hat{\text{EER}}^{\text{}}(\hat{\bar{Y}}_{d}^{\text{EBLUP}_{4}})/
\hat{\text{EER}}^{\text{}}(\hat{\bar{Y}}_{d}).$$
Como se puede observar en la Tabla 2 para todas las localidades la diferencia relativa es mayor a cero
y, por tanto se, disminuyen  los errores estándar relativos en todos los casos  al aplicar el modelo 4.

Las diferencias
m\'as notables se encuentran en las localidades con  tama\~nos de muestra peque\~nos: Usaquén,
Chapinero, Fontibón, Teusaquillo, Los mártires y Candelaria.
La Figura \ref{fig:mt4} muestra que en Usaquén y
Chapinero la diferencia relativa (entre los EER mediante
el estimador directo y el modelo Fay Herriot con interacci\'on) es considerable y por tanto
el uso del procedimiento propuesto es adecuado. Por otro lado,
es interesante observar como la precision aumenta para dichas localidades pero las estimaciones del ingreso promedio disminuyen
siendo una notable caracteristica para la toma de decisiones.

\clearpage

\section{Conclusiones}

Como producto de la investigación desarrollada en este trabajo se puede concluir:

\begin{itemize}

\item Se encuentra que mediante
la metodología propuesta, los errores estándar relativos en la estimación del ingreso promedio disminuyen.
\item Dichas estimaciones
son útiles no solo desde la perspectiva de la estimación en el área particular sino también en el planeamiento futuro de tamaños muestrales.
\item Se puede ser extendida para el cálculo de otros indicadores
\end{itemize}

Como trabajo futuro se podr\'ian explorar otro tipo de indicadores tales como la tasa de desempleo por localidad
en donde el problema de estimaci\'on, se convierte adem\'as, en un problema de estimaci\'on de dominios con
prevalencias peque\~nas en las \'areas seleccionadas.

\bibliographystyle{plainnat}

\bibliography{tesis}

\end{document}